\documentclass[aps,prb,twocolumn,showpacs,superscriptaddress]{revtex4}
\usepackage{graphicx}
\begin{document}

%\preprint{}
\title{Temperature dependence of the resonance and low energy spin excitations in superconducting 
FeTe$_{0.6}$Se$_{0.4}$}
\author{Leland W. Harriger}
\affiliation{
Department of Physics and Astronomy, The University of Tennessee, Knoxville, Tennessee 37996-1200, USA
}
\author{O. J. Lipscombe}
\affiliation{
Department of Physics and Astronomy, The University of Tennessee, Knoxville, Tennessee 37996-1200, USA
}
\author{Chenglin Zhang}
\affiliation{
Department of Physics and Astronomy, The University of Tennessee, Knoxville, Tennessee 37996-1200, USA
}
\author{Huiqian Luo}
\affiliation{Beijing National Laboratory for Condensed Matter Physics, Institute of Physics, Chinese Academy of Sciences, Beijing 100190, China
}
\author{Meng Wang}
\affiliation{Beijing National Laboratory for Condensed Matter Physics, Institute of Physics, Chinese Academy of Sciences, Beijing 100190, China
}
\author{Karol Marty}
\affiliation{Neutron Scattering Science Division, Oak Ridge National Laboratory, Oak
Ridge, Tennessee 37831-6393, USA}
\author{M. D. Lumsden}
\affiliation{Neutron Scattering Science Division, Oak Ridge National Laboratory, Oak
Ridge, Tennessee 37831-6393, USA}
\author{Pengcheng Dai}
\email{pdai@utk.edu}
\affiliation{
Department of Physics and Astronomy, The University of Tennessee, Knoxville, Tennessee 37996-1200, USA
}
\affiliation{Beijing National Laboratory for Condensed Matter Physics, Institute of Physics, Chinese Academy of Sciences, Beijing 100190, China
}

\begin{abstract}
We use inelastic neutron scattering to study the temperature dependence of the low-energy spin excitations in single crystals of superconducting
FeTe$_{0.6}$Se$_{0.4}$ ($T_c=14$ K). In the low-temperature superconducting state, the imaginary part of the dynamic susceptibility
 at the electron and hole Fermi surfaces nesting wave vector $Q=(0.5,0.5)$, $\chi^{\prime\prime}(Q,\omega)$, 
has a small spin gap, a two-dimensional neutron spin resonance above the spin gap, and increases  linearly with increasing $\hbar\omega$ for
energies above the resonance. While the intensity of the resonance decreases like an order parameter
 with increasing temperature and disappears at temperature slightly above $T_c$, the energy of the mode is weakly 
 temperature dependent and vanishes concurrently above $T_c$.  This suggests that in spite of its similarities with the resonance in electron-doped superconducting BaFe$_{2-x}$(Co,Ni)$_x$As$_2$, the mode in 
FeTe$_{0.6}$Se$_{0.4}$ is not directly associated with the superconducting electronic gap.
\end{abstract}


\maketitle

\section{Introduction}

Soon after the discovery of high temperature (high-$T_c$) superconductivity in Fe-based materials \cite{kamihara,mawkuen2,mawkuen1,fangmh}, 
neutron scattering studies revealed that 
the parent compounds of these superconductors have an antiferromagnetic 
(AFM) ground state \cite{cruz,wei,slli2} similar to those of unconventional 
heavy Fermions and copper oxide superconductors \cite{uemura}. This observation has 
inspired many theories to postulate that spin fluctuations in these materials may be responsible for
electron pairing and superconductivity \cite{mazin,chubkov,fwang,cvetkovic,moreo,mazin1,maier1,maier2,korshunov,seo09}. In one of the leading 
theories, superconductivity arises from quasiparticle excitations between the electron and hole pockets near $M$ and $\Gamma$ points of the Brillouin zone, respectively. One of the consequences of opening up electronic gaps in the superconducting state is that there should be a neutron spin resonance.
The energy of the resonance should be coupled to the addition of the hole and electron superconducting
gap energies ($\hbar\omega=\left|\Delta(k+Q)\right|+\left|\Delta(k)\right|$), and the intensity of the mode should
follow the superconducting order parameter \cite{maier1,maier2,korshunov}.  Indeed, the discovery of the neutron
spin resonance in electron and hole-doped iron pnictide BaFe$_2$As$_2$
at the AFM wave vector $Q=(0.5,0.5,L)$ in the tetragonal unit cell notation ($a=b=3.963$ and $c=12.77 $ \AA)  
\cite{christianson,lumsden,chi,pratt10,jtpark,mywang,clzhang,castellan} suggests that 
superconductivity arises from quasiparticle excitations between the signed reversed electron and hole pockets. This notion is further confirmed by the temperature \cite{inosov} and magnetic field \cite{jzhao} dependence of the mode energy which is directly coupled to the 
superconducting electronic gap energy . For iron chalcogenide Fe$_{1+\delta}$Te$_{1-x}$Se$_x$, previous inelastic
neutron scattering experiments \cite{mook,qiu,lumsden2,shlee,dnargyriou,slli10,zjxu,tjliu,wen2010,slli11,Igor,sxchi} have also established the presence of a resonance at the electron-hole Fermi surface 
nesting wave vector, which is the same as the AFM ordering wave vector for iron pnictides \cite{lumsden,chi,pratt10,jtpark,mywang,clzhang}, and the intensity of the resonance 
increases below $T_c$ just like it does for iron pnictides.  
Therefore, it appears that the neutron spin resonance is ubiquitous for different families of iron-based superconductors and directly correlated with superconducting electronic gaps \cite{dsinosov}.

In this article, we report inelastic neutron scattering studies of superconducting  FeTe$_{0.6}$Se$_{0.4}$ ($T_c=14$ K).   
Although there are extensive neutron scattering measurements on nonsuperconducting and superconducting 
Fe$_{1+\delta}$Te$_{1-x}$Se$_x$ \cite{mook,qiu,lumsden2,shlee,dnargyriou,slli10,zjxu,tjliu,wen2010,slli11,Igor,sxchi}, our detailed wave vector and energy dependent studies of the neutron spin resonance 
provide new information concerning the nature of the mode and its relationship to the superconducting electronic gap.
First, we confirm the earlier work \cite{qiu} that the mode is purely two-dimensional and dispersionless for wave vectors along the $c$-axis,
which is different from the dispersive nature of the resonance in electron-doped BaFe$_{2-x}$(Co,Ni)$_x$As$_2$ \cite{chi,pratt10,jtpark}. 
Second, we extend the earlier work \cite{qiu} on the temperature dependence of the mode. By carrying out systematic series of energy scans very close and above the superconducting transition temperature $T_c$, we find that the energy of the mode is essentially temperature independent and collapses at a temperature slightly above $T_c$, and does not follow the temperature dependence of the superconducting electronic gap as determined from Andreev reflection measurements \cite{wkpark}.  
Finally, we show that the intensity gain of the resonance is approximately compensated by spectral weight loss at energies below it, and there is a spin gap opening for low-energy spin excitations below $T_c$.  These results suggest that the neutron spin resonance in the FeTe$_{0.6}$Se$_{0.4}$ system may not be directly coupled to the superconducting electronic gap as those for BaFe$_{2-x}$(Co,Ni)$_x$As$_2$ \cite{inosov,jzhao}.  We discuss possible microscopic origins for this phenomenon.

\begin{figure}[t]
\includegraphics[scale=.4]{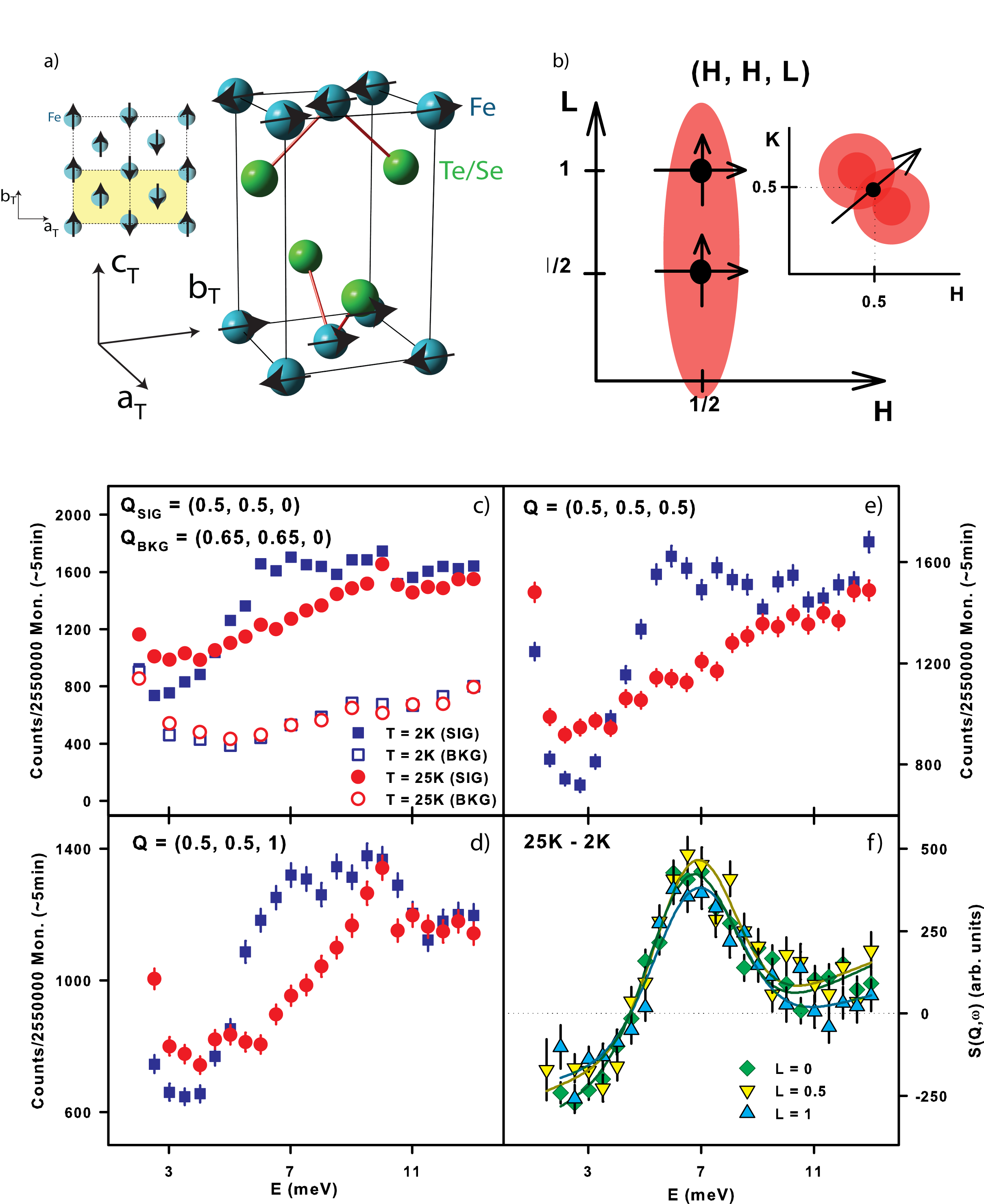}
\caption{
(color online). (a) Diagram of the Fe spin ordering with the shaded region defining the magnetic unit cell.  (b) Cartoon of the scan directions though the $(1/2, 1/2, L)$ nesting vector. The inset illustrates the direction in the $[H, K]$ plane that scans were confined to. Excitations at $(1/2, 1/2, L)$ in FeTe$_{1-x}$Se$_{x}$ consist of two incommensurate peaks that spread away from one another in the transverse direction. The red circles in the inset depict these excitations with the radius of the circles equal to twice the FWHM of the (1/2, 1/2, 0), 7.5 meV resonance peaks measured on crystals from the same batch on a different experiment. The separation of their centers is set to agree with the dispersion mapped out in this previous experiment \cite{slli10}.  (c-e) Energy scans about the 7 meV resonance position above and below $T_{c}$ for $L = 0, 1/2, 1$. Clear intensity gain is observed inside the superconducting state. The background at $L = 0$ is plotted above and below $T_{c}$ and is found to be identical, allowing direct temperature subtraction of the scans with no need for background correction.  (f) Temperature subtraction of energy scans shown in panels (c-e) demonstrating no observable dispersion of the resonance energy along $L$.
 }
\end{figure}

\section{Experimental Details}

We carried out neutron scattering experiments on the HB-3 thermal triple axis spectrometer at the High Flux Isotope Reactor (HFIR), Oak Ridge National Laboratory. We used a pyrolytic graphite PG(002) monochromator and analyzer with a collimation of 48$^{\prime\prime}$-monochromator-60$^{\prime\prime}$-sample-80$^{\prime\prime}$-analyzer-240$^{\prime\prime}$-detector. The data were collected in fixed E$_{f}$ mode at 14.7 meV with a PG filter placed between the sample and analyzer to remove contamination from higher order reflections. We coaligned two single crystals in the $[H, H, L]$ scattering plane and loaded them in a liquid He orange cryostat. The total mass was $\sim10$ grams with an in-plane and out-of-plane mosaic of 2.0$^{\circ}$ and 2.1$^{\circ}$ full-width at half maximum (FWHM), respectively. We defined the wave vector $Q$ at $(q_x, q_y, q_z)$ as $(H, K, L) = (q_xa/2\pi, q_yb/2\pi, q_zc/2\pi)$ reciprocal lattice units (rlu) using the tetragonal unit cell (space group P4/nmm), where $a = 3.8$ \AA, $b = 3.8$ \AA, and $c = 6.0$ \AA.
In the parent compound, FeTe, the AFM Bragg peaks occur at the $(1/2, 0, 1/2)$ and equivalent wave vectors, corresponding to the crystallographic spin arrangement depicted in Fig. 1a \cite{wei,slli2}. In the nonsuperconducting FeTe$_{1-x}$Se$_{x}$ samples ($x\leq 0.3$), spin excitations coexist at both the $(1/2, 0, 1/2)$ AFM wave vector, and the $(1/2, 1/2, L)$ wave vector associated with 
nesting of electron and hole pockets on the Fermi surface \cite{zjxu,tjliu,slli11,Igor,sxchi}. Upon reaching optimal doping, spin excitations at the AFM wave vector are suppressed, however, they remain strong near the nesting vector and consist of a commensurate resonance mode (in the superconducting state) sitting on top of an incommensurate magnetic signal that follows an hourglass dispersion at low energies \cite{slli10}. We chose the $[H, H, L]$ scattering plane for our experiments since this zone gives us full freedom to probe the $L$ dependence of the resonance. In general, the excitations in this system are extremely diffuse and, as a result, much broader than the instrumental resolution. To quantify this, we have calculated the resolution along the $(H, 1-H)$ direction at the $(0.5, 0.5)$ position as a function of energy. The resulting instrumental resolution width in FWHM is roughly 20 times smaller than the incommensurate peak separation. Thus our data collection is a good measure of signal centered directly at the $(0.5, 0.5)$ position. 

\section{Results}
In previous work on electron-doped BaFe$_{2-x}$(Co,Ni)$_x$As$_2$ superconductors, the neutron spin resonance has been found to be dispersive along the $c$-axis, occurring at slightly different energies for $L=0$ and $L=1$ \cite{chi,pratt10,jtpark,mywang}.  Although previous measurements suggest that the resonance in FeTe$_{1-x}$Se$_{x}$ is two-dimensional \cite{mook,qiu}, there have been no explicit measurements of the resonance at different $L$- values.  With this in mind, we have carried out detailed energy scans of bulk superconducting FeTe$_{0.6}$Se$_{0.4}$ at the resonance wave vector $(1/2, 1/2, L)$ as a function of temperature and $L$.  Figures 1c-1e show constant-$Q$ scans at the signal 
$Q=(0.5,0.5,0)$, $(0.5,0.5,0.5)$, $(0.5,0.5,1)$ and background $Q=(0.65,0.65,0)$ positions above and below $T_c$.
Consistent with earlier results \cite{mook,qiu}, we see a clear enhancement of scattering around $E\approx 7$ meV below $T_c$ at the
signal wave vectors for all the $L$ values probed. Figure 1f over-plots the temperature differences between 2 K and 25 K data for three
$L$ values.  It is clear that for all $L$ values the resonance energy is the same within the errors of our measurements ($E=6.95\pm 0.5$ meV). Therefore, the mode is indeed two-dimensional and has no dispersion along the $c$-axis \cite{mook,qiu}.

\begin{figure}[t]
\includegraphics[scale=.4]{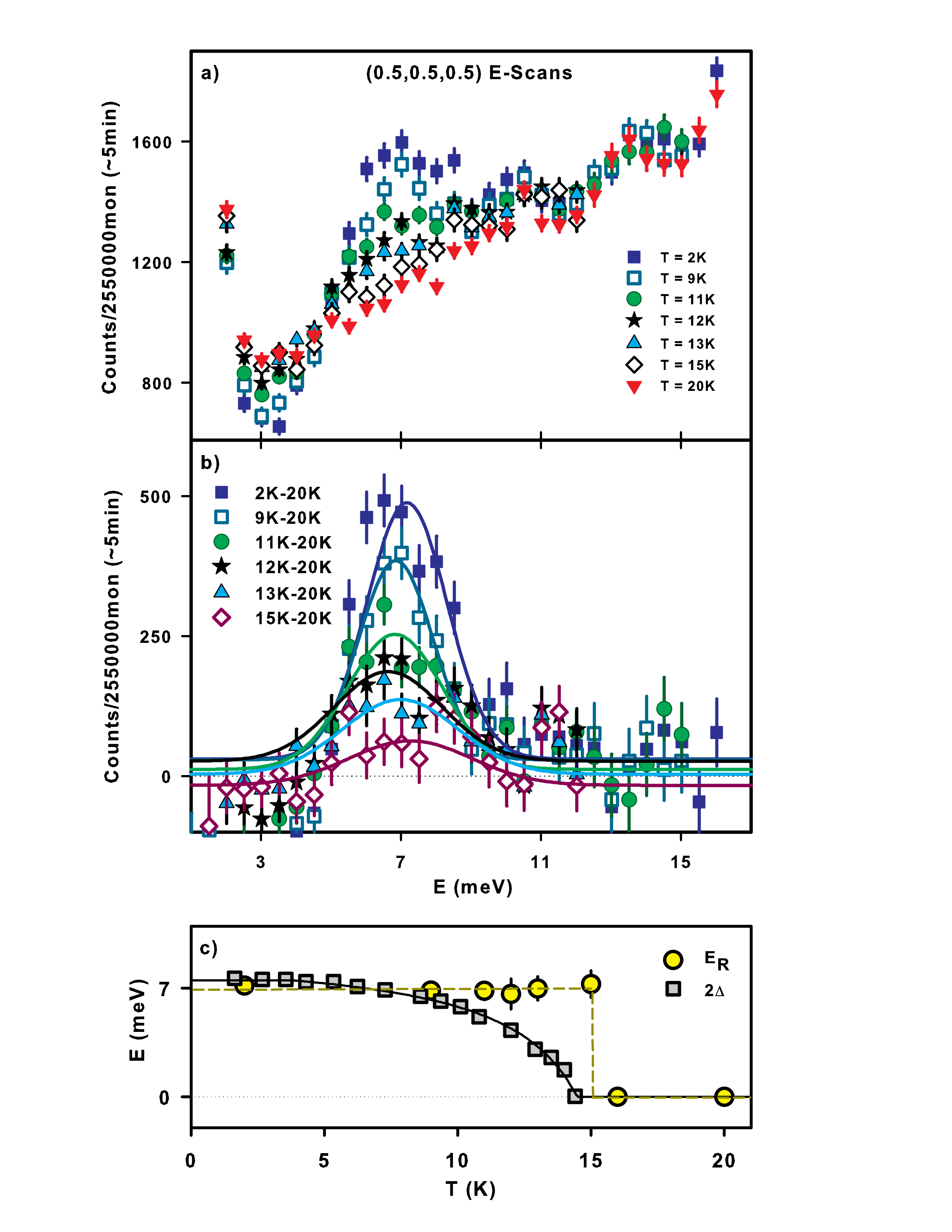}
\caption{
(color online). (a) Raw data for energy scans at $Q = (1/2,1/2,1/2)$ for multiple temperatures below $T_{c}$. At 2 K the 7 meV resonance is clearly present. A strong reduction in scattering for energies below 4 meV is also visible, indicating the opening of a gap in the system. Subsequent $Q$-scans, however, show that this is not a true gap. As the temperature increases to $T_{c}$ the resonance suppresses and the partial gap closes up.
(b) Temperature subtraction of scans shown in panel (a). All of the data is fit with a Gaussian leaving the center energy as a free parameter to be determined. (c) Position of the resonance energy vs temperature as determined from the fits in panel b), note that circle above $T = 15$ K are meant to indicate that the resonance has been completely suppressed. The temperature dependence of the superconducting gap \cite{wkpark} is also graphed, explicitly demonstrating that the resonance does not shift in energy as a function of temperature so as to remain inside $2\Delta$ as required by the spin exciton scenario.
}
\end{figure}

In previous neutron scattering experiments on optimally electron-doped BaFe$_{2-x}$Co$_x$As$_2$, careful temperature dependence measurements revealed that the energy of the resonance with increasing temperature tracks the temperature dependence of the superconducting gap energy \cite{inosov}. These results, as well as the magnetic field effect of the resonance \cite{jzhao}, provided compelling evidence that the resonance energy is intimately associated with the superconducting electronic gap energies.  To see if the resonance in FeTe$_{0.6}$Se$_{0.4}$ behaves similarly, we 
 carried out a series of energy scans from base temperature (2 K) to just above $T_{c}$ (20 K) 
at $Q=(0.5,0.5,0.5)$ (Fig. 2a). As the temperature is increased, we see that the resonance drops monotonically in intensity. 
To accurately determine the temperature dependence of the mode, the energy scans in the superconducting state were subtracted from the energy scan at 20 K in the normal state. The resulting plots of the resonance intensity gain were then fit to a Gaussian on a linear background with the center left as a free parameter (Fig 2b). 
By plotting the fitted values of the resonance energy as a function of temperature (Fig 2c),
we see that the resonance energy is essentially temperature independent until it abruptly disappears above $T_c$.
This is clearly different from the temperature dependence of the resonance for electron-doped BaFe$_{2-x}$Co$_x$As$_2$ \cite{inosov} and
the temperature dependence of the superconducting gap for FeTe$_{0.6}$Se$_{0.4}$ as determined from the Andreev reflection measurements (Fig. 2c) \cite{wkpark}.

\begin{figure}[t]
\includegraphics[scale=.5]{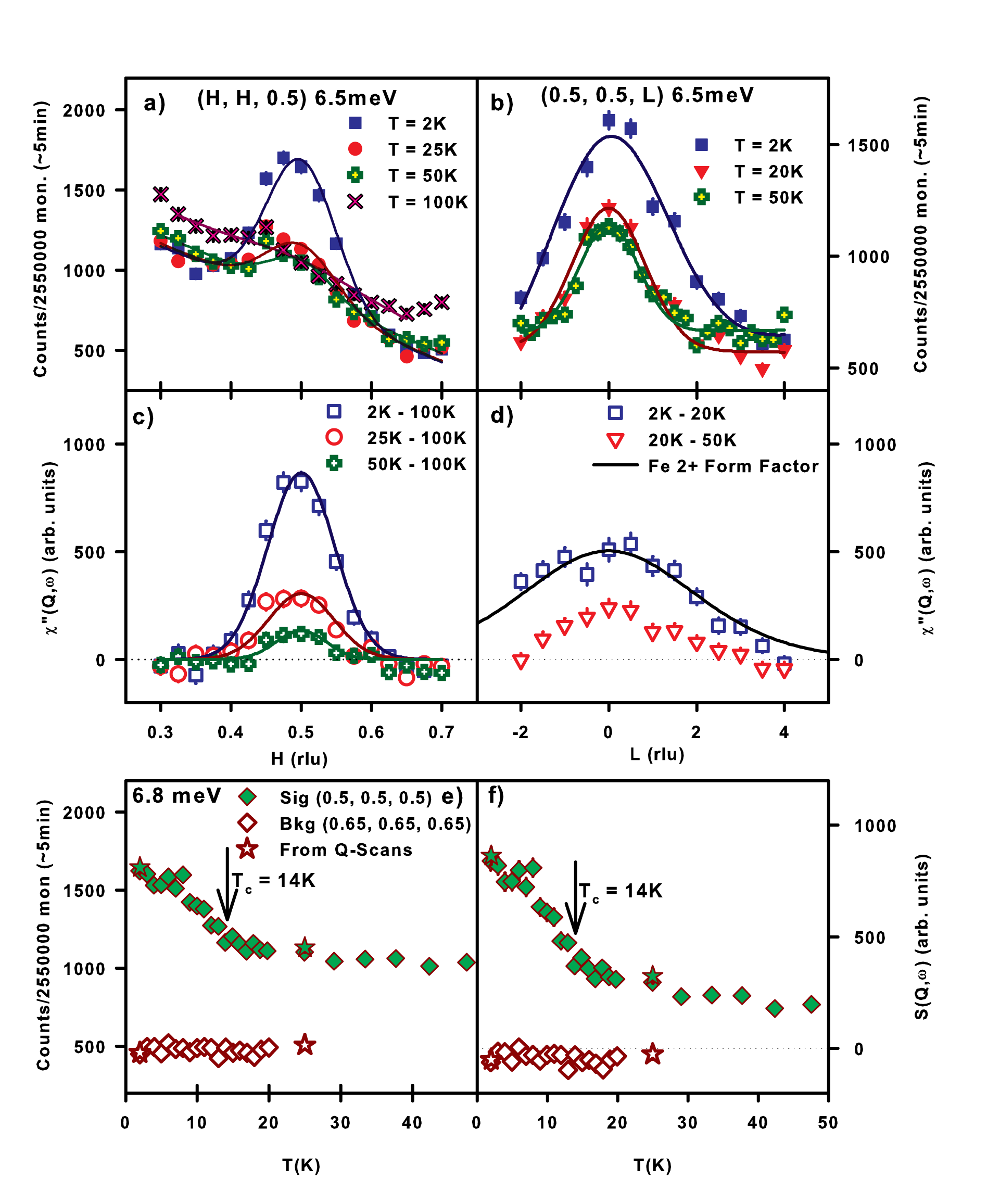}
\caption{
(color online.) (a,b) Raw $Q$-scan data along $[H, H]$ and $L$ respectively at $E_{R} = 6.5$meV at several temperatures. (c,d) $\chi^{\prime\prime}(Q,\omega)$ is determined by subtraction of the background and correcting for the Bose factor. In c) the 100 K data was used as a final background subtraction in order to remove a spurion at (0.45, 0.45, 0.5) and a phonon tail for points near (0.7, 0.7, 0.5). (d) The intensity gain due to the resonance is determined by subtraction of the 2 K and 20 K data. The resulting signal is very broad and fits well to the Fe$^{2+}$ form factor; a testament to the 2D nature of the resonant mode. (e,f) Temperature dependence of the resonance for $Q = (1/2, 1/2, 1/2)$ and $E = 6.8$ meV. The resonance suppresses as an order parameter as $T_{c}$ is approached.
 }
\end{figure}

\begin{figure}[t]
\includegraphics[scale=.4]{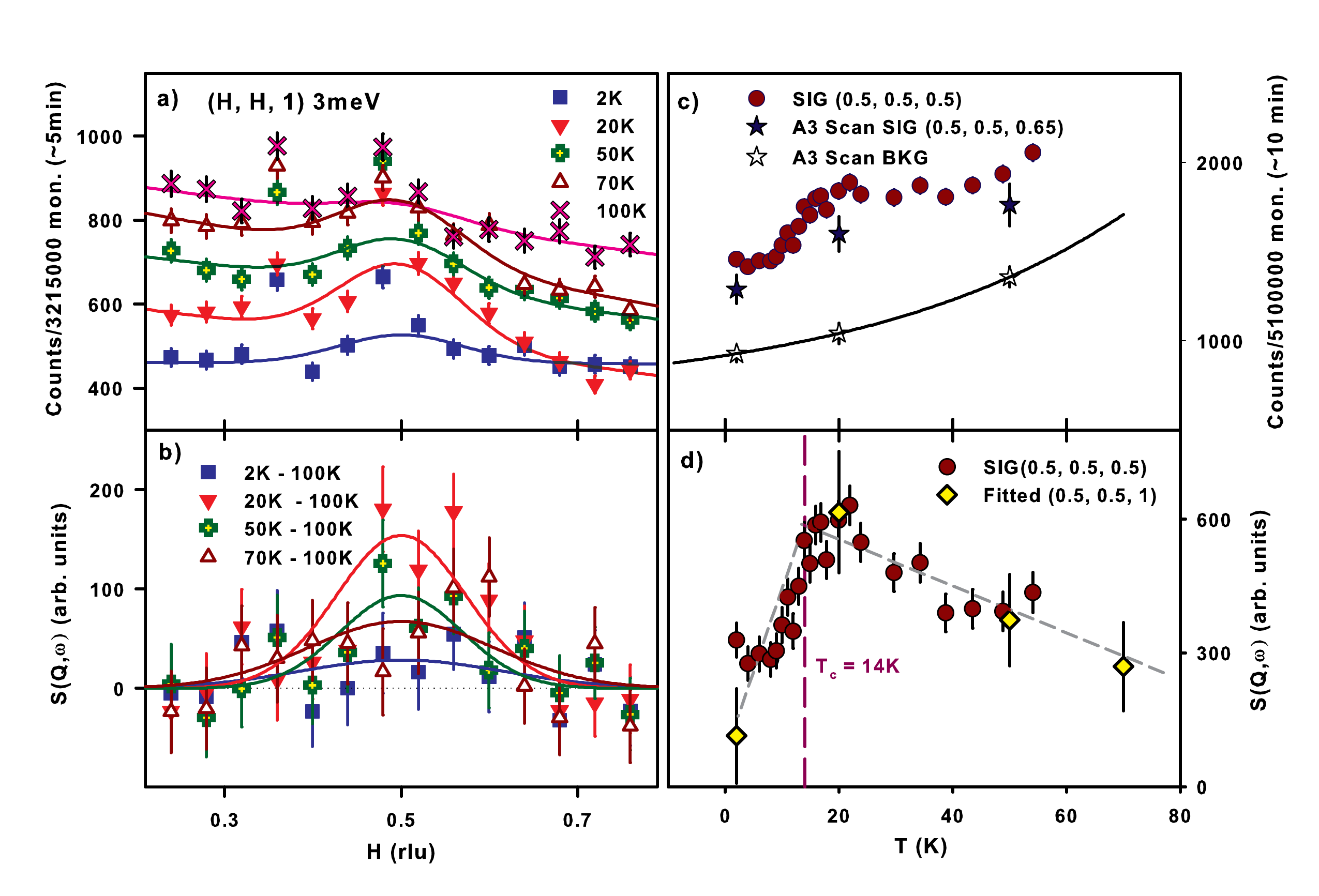}
\caption{(color online).
(a,b) $Q$-scan data along the $[H, H]$ direction for $L = 1$ and $E = 3$ meV. The scattering becomes stronger as $T_{c}$ is approached from higher temperatures, upon entering the superconducting state the intensity drops significantly by 2 K but does not fully gap. (c) Temperature dependence at 3 meV inside of the pseudo spin gap region reveals that near $T_{c}$ a gap begins to form but never fully forms by base temperature. (d) $S(Q,\omega)$ of the temperature scan as determined by interpolating and subtracting the background collected using A3 rocking curves. Yellow diamonds correspond to cross checks with fitted Q-scans from panels (a,b). Since the Q-scans and temperature scan were collected on different experiments, the data sets were not normalized to one another by monitor count but rather shifted to coincide at 20 K.
 }
\end{figure}

To further characterize the resonance, a series of $Q$-scans were carried out at $E = 6.5$ meV. 
Scans along the $[H, H]$ direction for $L = 0.5$ confirm that the resonance peaks at the $(0.5, 0.5)$ position with a strong gain in intensity in the superconducting state (Figs. 3a and 3c). 
For temperatures above 20 K, the drop in intensity is much more gradual with the peak at $(0.5, 0.5)$ fully suppressed by 100 K. Similar scans along the $[0.5,0.5,L]$ direction (Fig 3d-f) reveal that the scattering is much broader. The intensity gain of the resonance is extracted by subtraction of the 20 K and 2 K data. The $L$-dependence of the signal fits well to the Fe$^{2+}$ form factor, a further indication that the resonance is purely two-dimensional in nature. A temperature scan at $(0.5,0.5,0.5)$
for $E = 6.8$ meV confirmed that the resonance is strongest at base temperatures and then reduces like an order parameter to $T_{c}$ in good agreement with earlier measurements of the system \cite{mook,qiu,slli10,slli11}.

Interestingly, the 15 K energy scan in Fig 2b and the temperature dependence of the resonance in Figs. 3e and 3f suggest that the resonance mode first forms at a temperature slightly above $T_c$.  This behavior was also observed by Qiu {\it et al.} \cite{qiu} in their temperature and energy scans of the resonance in FeSe$_{0.4}$Te$_{0.6}$. A similar analysis on optimally electron doped 
BaFe$_{1.85}$Co$_{0.15}$As$_{2}$ \cite{inosov} and BaFe$_{1.9}$Co$_{0.1}$As$_{2}$ \cite{jzhao} does not display such behavior.  Although the origin of this effect is unclear, it is consistent with 
the idea of preformed Cooper pairs developing in the normal state just above $T_c$.  
For comparison, we note that the neutron spin resonance in underdoped YBa$_{2}$Cu$_{3}$O$_{6+x}$ can extend more than 50 K above $T_c$ \cite{dai1999}. 

From Figs. 1 and 2, we see that the intensity gain of the resonance in the superconducting state is accompanied by a loss in signal for energies below 4 meV, suggesting that conservation of spectral weight is satisfied by a reduction of scattering below the resonance energy. However, earlier measurements \cite{slli10} suggest
that the spin gap in FeTe$_{0.6}$Se$_{0.4}$ is unclean and does not fully open until $\sim$1 meV. 
Thus, it is interesting to investigate the temperature dependence of the spin excitations for energies above the spin gap and below the resonance. Figure 4a shows $Q$-scans along the $[H, H, 1]$ direction at different temperatures. 
With increasing temperature from 2 K, a peak at $(0.5,0.5,1)$ above 
background initially increases at $T=20$ K, then decreases upon further warming until disappearing at 100 K.  Assuming that there are only background scattering at 100 K, the temperature difference plots in Fig. 4b confirm that the magnetic scattering increases on warming to $T_c$ and then decreases with further increasing temperature.  Figure 4c shows the detailed temperature dependence data at 
the signal $Q=(0.5,0.5,0.5)$ and background (sample rotated away from the 
signal position by 30 degrees) position.  As we can see, the scattering 
shows a clear kink at $T_c$ and decreases monotonically above $T_c$ with warming.
Figure 4 shows the background corrected temperature dependence of the
magnetic scattering assuming that the temperature dependence of the
background follows the solid line in Fig. 4c.  
The effect of superconductivity is to open a pseudogap in spin excitations
spectrum below $T_c$.

For optimally electron BaFe$_2$As$_2$, the enhancement of the resonance occurs at the expense of a full spin gap opening below the resonance \cite{lumsden,chi}.  However, the situation for Fe(Se,Te) was not completely clear since there are no clean spin gaps for Fe(Se,Te).  Furthermore, it was not even clear whether the reduction in magnetic intensity at energies below the resonance occurs around $T_c$, when the resonance appears.  From our data, we see that this is indeed the case. It is worth noting that in lightly doped, nonsuperconducting FeTe, measurements at $(0.5, 0.5)$ also reveal a loss in scattering at 3meV. However, for this underdoped system no resonance is present to suck away spectral weight. Rather, the signal loss is due to the fact that at lower dopings there exists inelastic magnetic scattering at both $(0.5,0)$ and $(0.5, 0.5)$ with a strong crossover of spectral weight between these wave vectors occurring around 3 meV \cite{sxchi}.

\begin{figure}[t]
\includegraphics[scale=.45]{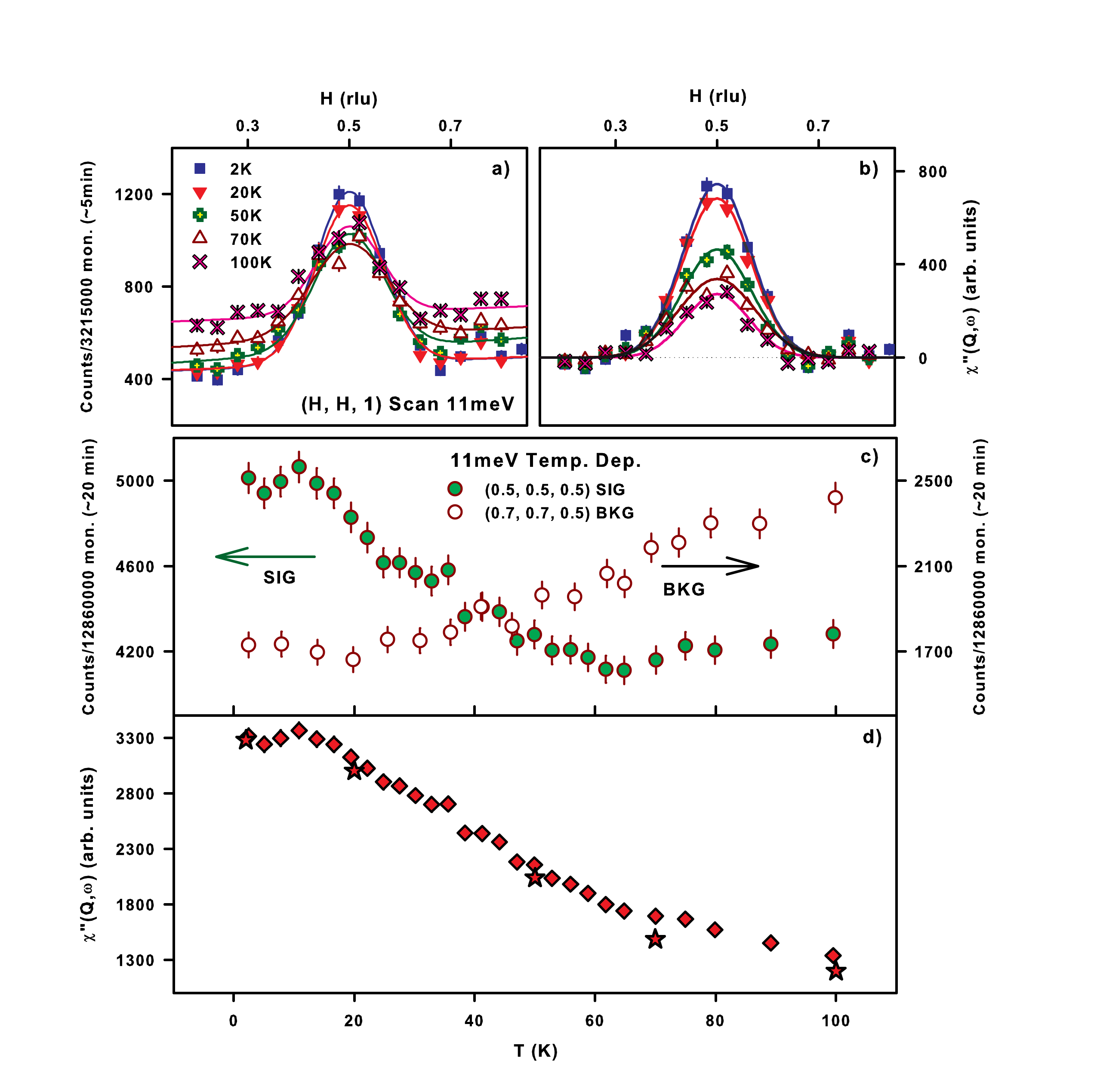}
\caption{(color online).
(a) Raw $Q$-scan data along the $[H, H]$ direction for $L = 1$ and $E = 11$ meV. (b) $\chi^{\prime\prime}(Q,\omega)$ determined by background subtraction and correcting for the Bose factor. The resonance is no longer visible, instead the scattering at 2 K is nearly identical to 20 K. Upon entering the normal state, the intensity begins dropping monotonically with increasing temperature but remains robust up to 100 K. (c,d) Temperature scan at (1/2, 1/2, 1/2) for $E = 11$ meV. Red stars correspond to cross checks with fitted peak intensities from $Q$-scans in panel a) that have been form factor corrected and normalized by monitor count.
 }
\end{figure}

\begin{figure}[t]
\includegraphics[scale=.65]{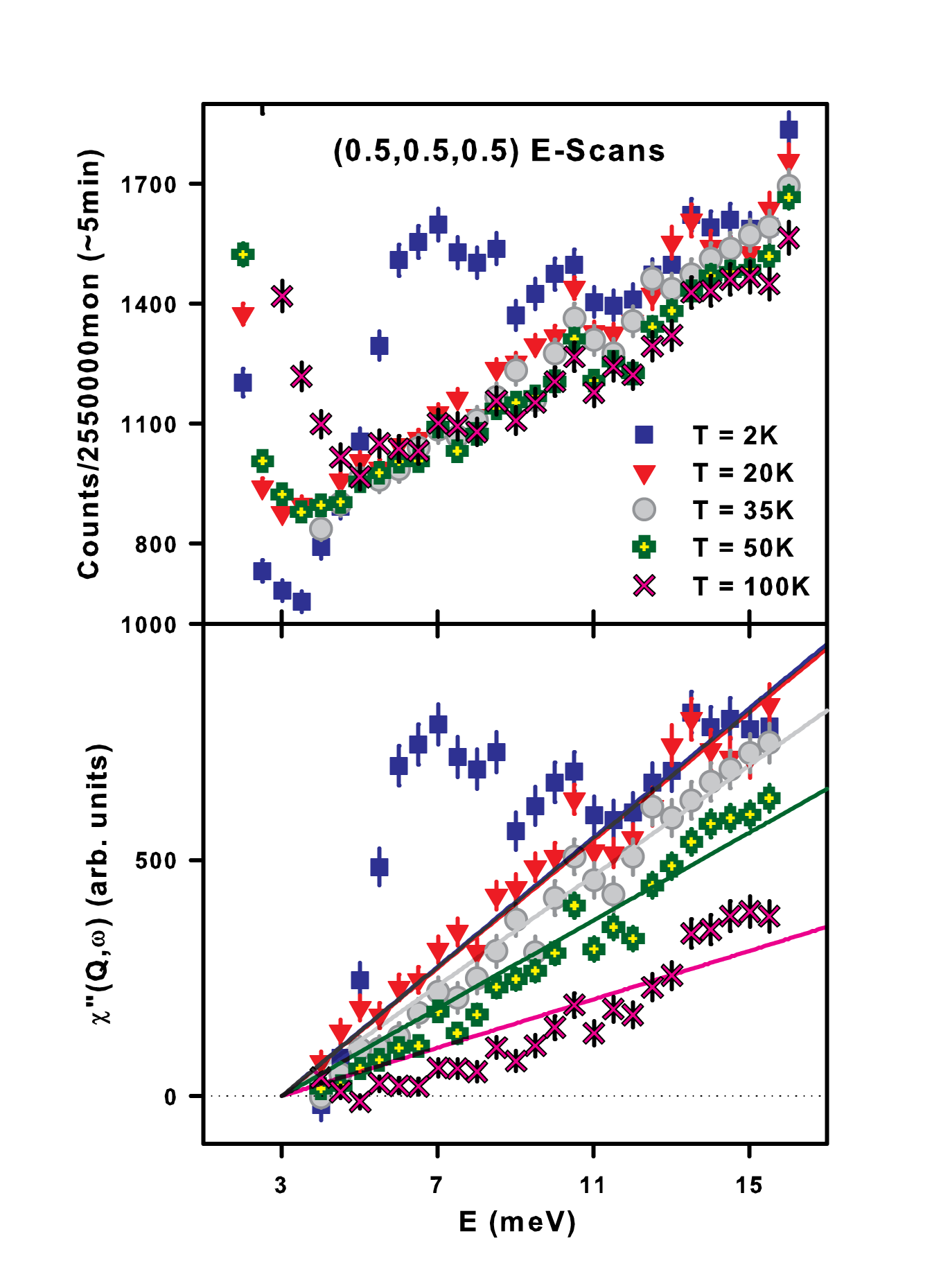}
\caption{(color online).
(a) Energy scans focusing on temperatures above $T_{c}$. (b) The background subtraction of $\chi^{\prime\prime}(Q,\omega)$ is determined from $Q$-scans. Aside from the resonance in the 2 K data, all other energy scans follow a similar linear trend; fanning out as a function of temperature.
 }
\end{figure}

To determine whether spin excitations at energies above the resonance also respond to superconductivity, we carried out a series of
constant-energy $E=11$ meV scans along the $[H,H,1]$ direction. The outcome shown in Figs. 5a and 5b 
reveals that magnetic scattering gradually increases in intensity on cooling. 
However, upon entering the superconducting state, the scattering appears to level off with the 2 K and 20 K $Q$-scans nearly identical in intensity. Temperature scans at $E=11$ meV at the signal [$Q=(0.5,0.5,0.5)$] and background [$Q=(0.7,0.7,0.5)$] 
positions are shown in Fig. 5c.  The background and Bose factor corrected temperature dependent imaginary part of the dynamic susceptibility, $\chi^{\prime\prime}(Q,E)$, at $Q=(0.5,0.5,0.5)$ and $E=11$ meV is shown in Fig. 5d. It is clear that the magnetic scattering grows with decreasing temperature but essentially saturates at temperatures below $\sim$15 K.

Finally, Figure 6a
shows the temperature evolution of the constant-$Q$ [
$Q=(0.5,0.5,0.5)$] scans from 2 K to 100 K. After correcting for the temperature 
dependence of the background scattering and Bose population factor, we obtain the
temperature dependence of $\chi^{\prime\prime}(Q,E)$ at $Q=(0.5,0.5,0.5)$ (Fig. 6b).
The $\chi^{\prime\prime}(Q,E)$ increases linearly with increasing energy,
and the resonance appears below $T_c$ together with the opening of a spin gap at lower energies.
These results are consistent with earlier work \cite{mook,qiu}.

\section{Discussions and Conclusions}

The presence of a neutron spin resonance in various high-$T_c$ copper oxide and Fe-based superconductors
has been suggested as the result of a spin-fluctuation mediated electron pairing mechanism \cite{dsinosov,gyu}.
In an earlier work mostly on copper oxide superconductors \cite{gyu}, it was proposed that the resonance energy is universally associated 
with the superconducting electronic gap $\Delta$ via $\hbar\omega_{res}/2\Delta=0.64$ instead of being proportional to the superconducting
transition temperatures $T_c$ \cite{wilson}.  
In a more recent summary of neutron scattering data on 
iron-based superconductors \cite{dsinosov}, it was found that
the energies of the resonance for underdoped BaFe$_{2-x}$(Co,Ni)$_x$As$_2$ deviate from this relationship, particularly for the resonance energy at $L=0$. 
For FeTe$_{0.6}$Se$_{0.4}$, angle resolved photoemission spectroscopy experiments 
\cite{hmiao} reveal a 4.2 meV gap on the electron Fermi surface and a 2.5 meV gap on the hole Fermi surface.  Since the addition
of the electron and hole superconducting electronic gap energies is consistent with the energy of the resonance at low temperature, the result has been interpreted as evidence that the resonance in FeTe$_{0.6}$Se$_{0.4}$ also arises from electron-hole pocket 
excitations \cite{hmiao}.  However, given that the superconducting gap energy gradually decreases for temperatures approaching $T_c$, the resonance energy will exceed that of the superconducting gap energy, contrary to the expectation for a spin exciton in the sign revised $s$-wave electron pairing scenario \cite{maier1,maier2,korshunov}.  

If superconductivity in iron-based materials is mediated by orbital
fluctuations associated with a fully gapped $s$-wave state without 
sign reversal ($s^{++}$-wave state), one would expect a neutron spin resonance at an energy above
the addition of the electron and hole superconducting electronic gap energies \cite{onari}.  Since the superconducting gaps decrease with increasing temperature, one would expect a reduction in the resonance energy with increasing temperature even in this scenario, contrary to the observation.  In the SO(5) theory for high-$T_c$
superconductivity \cite{demler}, the neutron spin resonance is a product of particle-particle excitations and is fixed in energy in the superconducting state.  Although this is consistent with present work, it remains unclear how the SO(5) theory originally designed for high-$T_c$ copper oxide superconductors would apply in the case of iron-based superconductors.  In any case, our results suggest that the resonance itself may not be directly associated with the superconducting electronic gap in the FeTe$_{0.6}$Se$_{0.4}$ system.

\section{ACKNOWLEDGMENTS}
We thanks Jiangpin Hu for helpful discussions.
This work is supported by the U.S. NSF-DMR-1063866.  
The single crystal growth efforts at UTK is supported by U.S. DOE BES No. DE-FG02-05ER46202. 
Work at IOP is supported by the Chinese
Academy of Sciences and by the Ministry of Science and Technology of China 973
program (2012CB821400, 2011CBA00110).
HFIR is supported by the US
Department of Energy, Division of Scientific User Facilities, Basic Energy Sciences.

%\bibliography{NoEndingPoint}

\end{document}